\documentclass{aa}  

\usepackage{graphicx}
\usepackage{lscape}
\usepackage{txfonts}
\usepackage[]{natbib}
\usepackage{enumitem}
\usepackage{multirow}
\begin{document}

   \title{XMMFITCAT: The XMM-Newton spectral-fit database\thanks{Based on observations obtained with XMM-Newton, an ESA science mission with instruments and contributions directly funded by ESA Member States and NASA}}

  \author{A. Corral
          \inst{1}
          \and
          I. Georgantopoulos\inst{1}
\and
M.G. Watson\inst{2}
\and
S.R. Rosen\inst{2}
\and
K.L. Page\inst{2}
\and
N.A. Webb\inst{3}
          }

   \institute{IAASARS, National Observatory of Athens, GR-15236 Penteli, Greece\\
              \email{acorral@noa.gr}
\and
Department of Physics \& Astronomy, University of Leicester, Leicester, LE1 7RH, UK
\and Institut de Recherche en Astrophysique et Planétologie (IRAP), 9 Avenue du Colonel Roche, 31028 Toulouse Cedex 4, France
             }

   \date{Received , ; accepted , }

 
  \abstract{The XMM-Newton spectral-fit database (XMMFITCAT) is a
    catalogue of spectral fitting results for the source detections
    within the {\it XMM-Newton Serendipitous} source catalogue with
    more than 50 net (background-subtracted) counts per detector in
    the 0.5-10 keV energy band. Its most recent version, constructed
    from the latest version of the {\it XMM-Newton} catalogue, the
    3XMM {\it Data Release} 4 (3XMM-DR4), contains spectral-fitting
    results for $\gtrsim$ 114,000 detections, corresponding to
    $\simeq$ 78,000 unique sources. Three energy bands are defined
    and used in the construction of XMMFITCAT: Soft (0.5-2 keV), Hard
    (2-10 keV), and Full (0.5-10 keV) bands. Six spectral models,
    three simple and three more complex models, were implemented and
    applied to the spectral data. Simple models are applied to all
    sources, whereas complex models are applied to observations with
    more than 500 counts (30\%). XMMFITCAT includes best-fit
    parameters and errors, fluxes, and goodness of fit estimates for
    all fitted models. XMMFITCAT has been conceived to provide the
    astronomical community with a tool to construct large and
    representative samples of X-ray sources by allowing source
    selection according to spectral properties, as well as
    characterise the X-ray properties of samples selected in different
    wavelengths. We present in this paper the main details of the
    construction of this database, and summarise its main
    characteristics.}

\titlerunning{XMMFITCAT: The XMM-Newton spectral-fit database}
\authorrunning{A. Corral et al.}
\maketitle

%

\section{Introduction}

X-rays observations have expanded our knowledge of the most energetic
phenomena in the Universe, and the astronomical sources in which they
take place, such as stars, galaxies, clusters, and active galactic
nuclei (AGN). Besides, and thanks to their penetrating ability, X-ray
observations detect sources hidden behind large amount of gas, and up
to high distances. Serendipitous X-ray surveys conducted by the {\it
  XMM-Newton}\footnote{http://xmm.esac.esa.int/} and {\it
  Chandra}\footnote{http://chandra.harvard.edu/} observatories, have
almost completely resolved the X-ray Cosmic Background (CXB) below 10
keV, showing that the X-ray sky is dominated by AGN emission
\citep{gilli07,treister09}.

Thanks to its large collecting area and its large field of view, {\it
  XMM-Newton} has proven to be an extraordinary instrument to perform
X-ray surveys. The European Photon Imaging Camera
(EPIC\footnote{http://xmm.esac.esa.int/external/xmm\_user\_support/documentation/\\technical/EPIC/index.shtml})
on-board {\it XMM-Newton} works in a photon-counting mode. This
characteristic, along with its large effective area, allows us, from a
single observation, to extract images, light curves, and spectral data
not only for the proposed target, but also for many of the detected
sources within the field of view. Data from this camera have been used
to construct the largest catalogue of X-ray sources ever built, the
{\it XMM-Newton} serendipitous source catalogue. The latest version of
the {\it XMM-Newton} catalogue, 3XMM Data Release 4
(3XMM-DR4\footnote{http://xmmssc-www.star.le.ac.uk/Catalogue/3XMM-DR4/}),
contains high-quality photometric information for more than 500~000
source detections, and light curves and spectral data for more than
120~000 of them.

The {\it XMM-Newton} spectral-fit database is an ESA funded project
aimed to provide the astronomical community with a tool to construct
large and representative samples of X-ray sources according to their spectral
properties. To this end, the database is built from the spectral data
within the {\it XMM-Newton} catalogue, which are then used to produce a new
catalogue composed of X-ray spectral fitting results.

\section{The XMM-Newton serendipitous source catalogue}
\label{3mmdr4}
The 3XMM-DR4 source catalogue is the largest catalogue of X-ray
sources compiled to date (see Rosen et al, in prep., and
\citealt{watson09}). It derives from 7427 pointed X-ray observations,
covering $\sim$2\% of the sky, taken with the EPIC camera pn and MOS
instruments on board the {\it XMM-Newton} satellite. The 3XMM-DR4
catalogue, which contains more than 500~000 detections from $\sim$
372~000 unique X-ray sources, was produced by the XMM-Newton Survey
Science Centre (SSC\footnote{http://xmmssc.irap.omp.eu/})
\citep{watson01} on behalf of ESA.\\

\subsection{Catalogue production}
Only data from the three EPIC (pn, MOS-1 and MOS-2) instruments are
used in the 3XMM-DR4 catalogue. These instruments operate in
photon-counting mode such that, from a single observation, spectra and
time series can be extracted from every source detected in the observed
field of view. During routine pipeline processing of the XMM-Newton
EPIC data, photometric measurements are obtained for every detected
source in a number of distinct energy bands over the 0.2-12 keV range
(see \citealt{watson09}). In parallel, for brighter sources ($>$ 100
counts, summed over all 3 instruments), spectra and time series are
extracted.

The 3XMM-DR4 catalogue reflects both an increased number of
detections, due to the increased (3.2 yr) observation baseline since
its predecessor, 2XMMi-DR3, and improvements to the processing system
and calibration information. The number of public observations
contained in 3XMM-DR4 increased by $\sim$ 50\% relative to
2XMMi-DR3. At the same time, the science data processing has taken
advantage of significant improvements within the XMM-Newton Science
Analysis Software (SAS) and calibration data. The key science-driven
gains include:

\begin{itemize}
\item{Improved source characterisation and reduced spurious source detections.}
\item{Improved astrometric precision of sources.} 
\item{Greater net sensitivity for source detection.}
\item{Extraction of spectra and time series for fainter sources, with
  improved signal-to-noise ratio (SNR).}
\end{itemize}

The resulting catalogue contains spectra for more than 120~000
detections corresponding to $\sim$ 80~000 unique sources.

\subsection{Spectral extraction}
The pipeline processing automatically extracts spectra and time series
(source-specific products, SSPs), from suitable exposures, for
detections that meet certain brightness criteria.\\

In previous versions of the processing pipeline, extractions were
attempted for any source which had at least 500 EPIC counts. In such
cases, source data were extracted from a circular aperture of fixed
radius (28 arcseconds), centred on the detection position, while
background data were accumulated from a co-centred annular region with
inner and outer radii of 60 and 180 arcseconds, respectively. Other
sources that lay within or overlapped the background region were
masked during the processing. In most cases this process worked
well. However, in some cases, especially when extracting SSPs from
sources within the small central window of MOS Small-Window mode
observations, the background region could comprise very little usable
background, with the bulk of the region lying in the gap between the
central CCD and the peripheral ones. This resulted in very small (or
even zero) areas for background rate scaling during background
subtraction, often leading to incorrect background subtraction during
the analysis of spectra in {\tt
  Xspec}\footnote{http://heasarc.gsfc.nasa.gov/xanadu/xspec/}, the
standard package for X-ray spectral analysis.\\

For the bulk reprocessing leading to 3XMM-DR4, two new approaches have
been adopted and implemented in the pipeline.
\begin{enumerate}
\item{The extraction of data for the source takes place from an
    aperture whose radius is chosen to maximise the signal-to-noise
    (SNR) of the source data. This is achieved by a curve-of-growth
    analysis, performed by the SAS task, {\tt eregionanalyse}. This is
    especially useful for fainter sources where the relative
    background level is high.}
 \item{To address the problem of locating an adequately filled
   background region for each source, the centre of a circular
   background aperture of radius, rb=168 arcseconds (comparable area
   to the previously used annulus) is stepped around the source along
   a circle centred on the source position. Up to 40 uniformly spaced
   azimuthal trials are tested along each circle. A suitable
   background region is found if, after masking out other
   contaminating sources and allowing for empty regions, a filling
   factor of at least 70\% usable area remains. If no background trial
   along a given circle yields sufficient residual background area,
   the aperture is moved out to a circle of larger radius and the
   azimuthal trials are repeated. The smallest trial circle has a
   radius of rc=rb + 60 arcseconds so that the inner edge of the
   background region is at least 60 arcseconds from the source centre
   - for the case of MOS Small-Window mode, the smallest test circle
   for a source in the central CCD is set to a radius that already
   lies on the peripheral CCDs. Other than for the MOS Small-Window
   cases, a further constraint is that, ideally, the background region
   should lie on the same instrument CCD as the source.}
\end{enumerate}

If no solution is found with at least a 70\% filling factor, the
background trial with the largest filling factor is adopted.\\

For the vast majority of detections where SSP extraction is attempted,
this process obtains a solution in the first radial step, and a strong
bias to early azimuthal steps, i.e. in most cases an acceptable
solution is found very rapidly. For detections in the MOS instruments,
about 1.7\% lie in the central window in Small-Window mode and have a
background region located on the peripheral CCDs. Importantly, in
contrast to earlier pipelines, this process always yields a usable
background spectrum for objects in the central window of MOS
Small-Window mode observations.\\

In addition, the current pipeline permits extraction of SSPs for
fainter sources. Extraction is considered for any detection with at
least 100 EPIC source counts in the full catalogue band (0.2-12 keV),
instead of the 500 EPIC counts limit used in the previous
pipeline. Where this condition is met, spectra and time series are
  extracted. For a more detailed description of the 3XMM-DR4 catalogue
  production see
  http://xmmssc-www.star.le.ac.uk/Catalogue/3XMM-DR4/UserGuide\_xmmcat.html. Catalogue
  spectra and time series can be retrieved from the {\it XMM-Newton}
  Science Archive (XSA\footnote{http://xmm.esac.esa.int/xsa/}), as
  well as from the web services listed at the end of
  Sect.~\ref{overview}.

\section{The XMM-Newton spectral-fit database}
\label{database}
The {\it XMM-Newton} spectral-fit database is a project aimed to take
advantage of the great wealth of data and information contained within
the {\it XMM-Newton} serendipitous source catalogue to construct a
database composed of spectral-fitting results. The possible
applications of this database include: the construction of large and
representative samples of X-ray sources according to spectral
properties; the possibility of pinpoint sources with interesting
spectral properties; and the cross-correlation with samples selected
in other wavelengths so as to obtain a first-order description of the
sources X-ray spectral-properties.

\subsection{Automated spectral fitting}

The {\it XMM-Newton} spectral-fit database (XMMFITCAT) is constructed
by using automated spectral fits applied to the pipeline extracted
spectra within the 3XMM-DR4 catalogue. The software used to perform
the spectral fits was {\tt Xspec} v12.7 \citep{arnaud}. In the case of
multiple observations available for the same source, each observation
is fitted separately. The fitting scripts were designed to jointly fit
all the available spectra for the same source detection. As a result,
the database contains spectral-fitting results for each source
observation, not for each observed source.

The data used in the construction of the database are spectral data
(source and background spectra), as well as ancillary matrices,
retrieved from the 3XMM-DR4 catalogue products. Redistribution
matrices are the canned matrices provided by the {\it XMM-Newton} SOC
(Science Operations Centre).

The spectral-fitting pipeline is composed of tcl (Tool Command
Language), and Perl scripts. The default fitting algorithm that {\tt
  Xspec} uses to find the best-fit values for each model parameter is
a modified Levenberg-Marquardt algorithm. This default algorithm is
the one used in the construction of the database so as to optimise the
fitting speed. However, this algorithm is local rather than global, so
it is possible for the fitting process not to find the global
best-fit, but a local minimum. To prevent this, the scripts include an
optimisation algorithm that tries to avoid the fit to fall into a
local minimum by computing the errors on all variable parameters
  at the 95\% confidence level. If a better fit is found, the
  optimisation algorithm starts again. In the case non-monotonicity is
  detected during the error computation, the confidence level in the
  error computation is increased until a better minimum is
  found. Once the best-fit is found, errors are computed and reported
  within the database at the 90\% confidence level.
\begin{table}[ht]
\caption{Energy bands}
\begin{center}
\begin{tabular}{|c|c|c|c|c|}
\hline
Energy range &\multicolumn{2}{c|}{3XMM-DR4} & \multicolumn{2}{c|}{XMMFITCAT} \\
(keV) &\multicolumn{2}{c|}{bands} & \multicolumn{2}{c|}{bands} \\
\hline
0.2 - 0.5 & 1 & \multirow{6}{*}{Total} & \multicolumn{2}{c|}{}\\
\cline{1-2}\cline{4-5} 
0.5 - 1.0 & 2 & & \multirow{2}{*}{Soft} &\multirow{4}{*}{Full}\\
\cline{1-2}
1.0 - 2.0 & 3 & & & \\
\cline{1-2}\cline{4-4}
2.0 - 4.5 & 4 & & \multirow{2}{*}{Hard} & \\
\cline{1-2} 
4.5 - 10 & \multirow{2}{*}{5} & & & \\
\cline{1-1}\cline{4-5}
10 - 12 & & & \multicolumn{2}{c|}{}\\
\hline
\end{tabular}
\end{center}
\label{bands}
\end{table}

The energy bands used in the automated fits are listed in
Table~\ref{bands}. As a comparison, the energy bands used in the
construction of the 3XMM-DR4 catalogue are also listed in the same
table.

\subsection{Spectral data selection}

Cash statistics, implemented as C-stat in {\tt Xspec}, are used to fit
the data. This statistic was selected, instead of the more commonly
used $\chi^2$ statistic, to optimise the spectral fitting in the case
of low count spectra. The 3XMM-DR4 spectra are unbinned and then
binned to 1 count/bin. The combined use of spectra binned to 1
count/bin plus C-stat fitting has been proven to work very well when
fitting spectral data down to 40 counts \citep{krumpe}.

During the spectral fits, all variable parameters for different
instruments and exposures are tied together except for a relative
normalisation, which accounts for the differences between different
flux calibrations. 
Given that each additional instrument spectrum adds a new parameter to
the fit, and to ensure a minimum quality on the spectral fits, a lower
limit on the number of counts in each individual spectrum is imposed:
only spectra corresponding to a single EPIC instrument, with more than
50 source counts in the {\it Full} band are included in the spectral
fits. Note that this implies that not all available spectra
  within 3XMM-DR4 for a given observation are used in some cases. A
  table listing the spectra used for each source observation is also
  publicly available, along with the one containing the spectral
  fitting results, at the database's webpage (see Sect.~\ref{overview}).

\subsection{Spectral models}

Three simple, and three more complex models have been implemented. All
these models are applied to the spectral data if the following
conditions are fulfilled : 

\begin{itemize}
\item Simple models: total number of counts (all instruments added
  together) larger than 50 counts in the energy band under
  consideration.  
\item Complex models: total number of counts larger than 500 counts in
  the {\it Full} band.
\end{itemize}

The models are selected to represent the most commonly observed
spectral shapes in astronomical sources in a phenomenological way. The
preferred model (among the implemented ones) for each observation is
selected according to the goodness of each fit (see
Sect.\ref{good}), but spectral-fitting results for all the
models applied are included in the database. It is important to note
that the automated procedure is only intended to obtain a good
representation of the spectral shape so, given the limited number of
spectral models applied, the preferred model should not be interpreted
as a ``best-fit model'' in the way it is when carrying out manual
fits.

The simple models (models 1 to 3), and the more complex models (models
4 to 6) are:

\begin{enumerate} 
\item {\bf Absorbed power-law model} ( {\tt wabs*pow} in {\tt Xspec}
  notation ): A power-law model modified by photoelectric absorption.
\item{\bf Absorbed thermal model} ({\tt wabs*mekal)}: A thermal model
  modified by photoelectric absorption.
\item{\bf Absorbed black-body model} ({\tt wabs*bb)}: A
  photoelectrically absorbed black-body model.
\item{\bf Absorbed power-law model plus thermal model} ({\tt
  wabs(mekal+wabs*pow)}): A thermal plus a power-law model in which
  both components are modified by absorption, and the power-law is
  additionally absorbed.
\item{\bf Double power-law model} ({\tt wabs*(pow+wabs*pow})): A
  double power-law model, with different photon indices, modify by
  photoelectric absorption, and additional absorption only affecting
  one of the power-law components.
\item{\bf Black-body plus power-law model} ({\tt wabs*(bb+pow)}): A
  black-body plus a power-law component modified by photoelectric
  absorption.
\end{enumerate}

\begin{figure*}[!htbp]
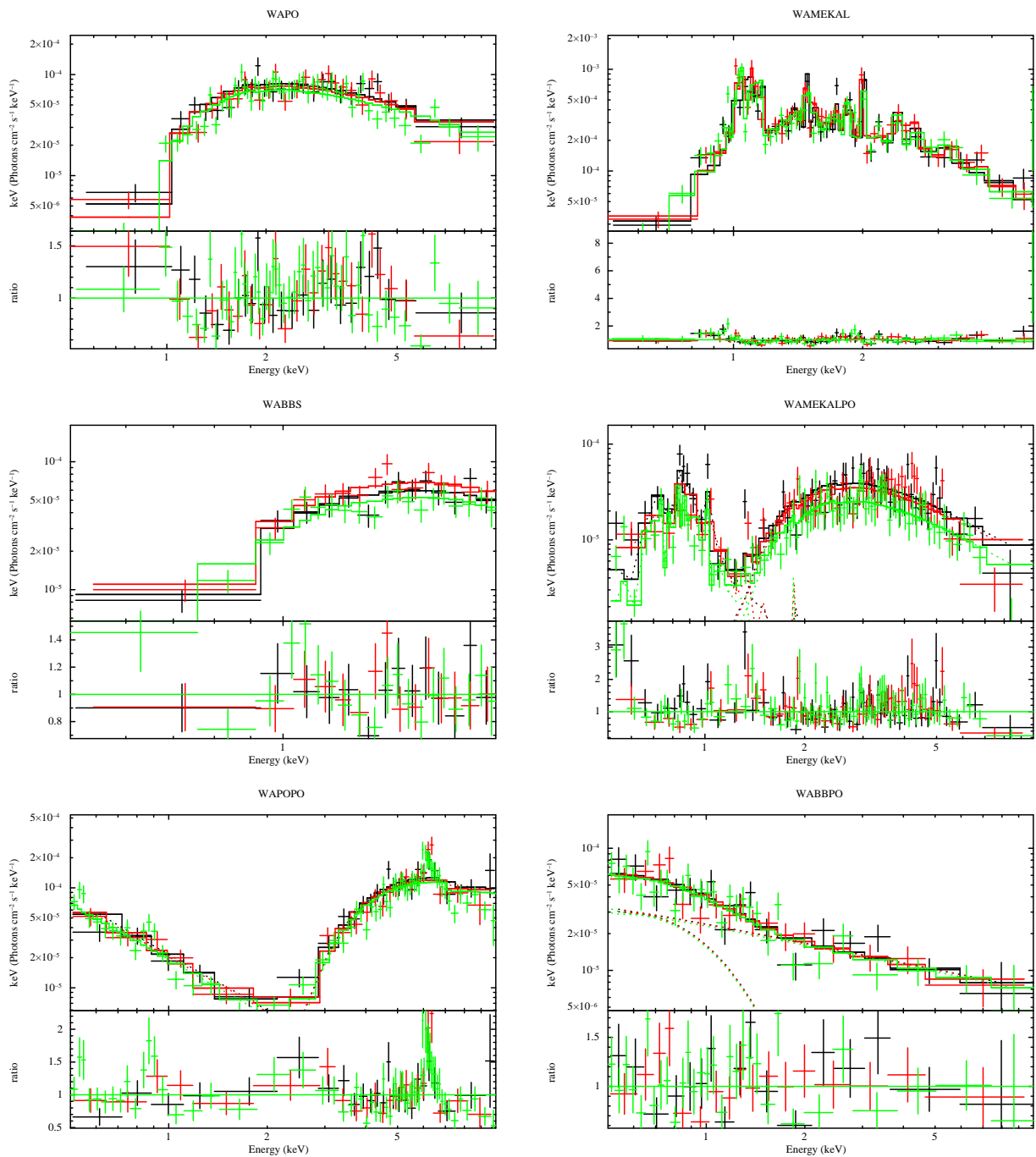

\centering
$$
\begin{array}{cc}
    \includegraphics[angle=-90,width=8cm]{figure1a.ps} &    \includegraphics[angle=-90,width=8cm]{figure1b.ps} \\
    \includegraphics[angle=-90,width=8cm]{figure1c.ps} &    \includegraphics[angle=-90,width=8cm]{figure1d.ps} \\
    \includegraphics[angle=-90,width=8cm]{figure1e.ps} &    \includegraphics[angle=-90,width=8cm]{figure1f.ps} \\
  \end{array}
$$
  \caption{Examples of spectral-fitting results for the models (from
    top to bottom and left to right): absorbed power-law model;
    absorbed thermal model; absorbed black-body model; thermal plus
    power-law model; double power-law model; and black-body plus
    power-law model.}
\label{samplemodels}
\end{figure*}

A component is considered not-significant in a complex-model fit if
its corresponding normalisation is consistent with zero at the 90\%
confidence level. In those cases, spectral-fitting results for that
model are not included in the database. Plots corresponding to each of
the implemented models are shown in Fig.~\ref{samplemodels}. Each
model and its corresponding spectral-fitting process are described in
Sects.\ref{po} to \ref{bbpo}. Parameters and allowed ranges for them
that are not described in these sections are set to their {\tt Xspec}
default values. The initial value and allowed range of values for the
column density for all absorption components ({\tt wabs}) is the same
for all simple models: initial value N$_{H}$ = 10$^{21}$ cm$^{-2}$;
allowed range: 10$^{20}$ to 10$^{24}$ cm$^{-2}$.

As part of the spectral fits, errors are computed at the 90\%
confidence level for every variable parameter. If {\tt Xspec} cannot
constrain the value of a certain parameter, i.e the error computation
pegs at the lower and upper limits of the allowed range, the parameter
is fixed. Fixed values are not the same for all observations, but they
are computed during each spectral fit (see below), and they depend on
the spectral model, energy band, and data quality.

\subsubsection{Absorbed power-law model}
\label{po}
A power-law component is the most common spectral shape displayed by
X-ray sources, including active galactic nuclei (AGN, the most
abundant X-ray sources), and X-ray binaries. This model is applied in
the {\it Full}, {\it Soft}, and {\it Hard} bands to get a first-order
characterisation of the full spectral shape. Besides, a simple
absorbed power-law model is found to be a good representation of most
low to medium quality AGN spectra. The variable parameters of the
model are the column density of the absorption component, and the
photon index (initial value $\Gamma$ = 2; range: 0-4), and
normalisation of the power-law component.

The first step of the spectral fit for this model in the {\it Full}
band is to compute the photon index in the {\it Hard} band without
including absorption. The resulting value is the one used to fix the
photon index of the model if it cannot be constrained. Sometimes the
number of counts above 2 keV is $<$ 20 counts, and this value cannot
be computed. In those cases, the fixed value is the initial value,
$\Gamma$ = 2. If the parameter that cannot be constrained is the
column density, its fixed value is the one obtained by carrying out
the spectral fit in the {\it Soft} band with $\Gamma$ fixed to the
value obtained in the {\it Hard} band. If the number of counts below 2
keV is $<$ 20 counts, the fixed value is N$_{H}$ = 10$^{22}$
cm$^{-2}$. This is a reasonable approximation since spectra with $>$
50 counts in the {\it Full} band but $<$ 20 counts below 2 keV are
most likely absorbed by column densities above that value.

The fixed values for the photon index and the column density when
fitting in the {\it Soft} and {\it Hard} bands are $\Gamma$=2 and
N$_{H}$ =10$^{22}$ cm$^{-2}$, respectively. We adopted these values
because it is very difficult to constrain both the photon index and
the column density in the {\it Soft} band in the case of low-count
spectra, and the spectral fit is insensitive to column densities below
that value in the {\it Hard} band. The spectral-fitting results of
this model applied in these narrow bands are used as initial
parameters and fixed values in the complex models.

\subsubsection{Absorbed Thermal model}
The absorbed thermal model is applied in the {\it Full} and {\it Soft}
bands. It is intended to model emission from stars, galaxies, and
galaxy clusters. The {\tt Xspec mekal} model is used, instead of
  the more up to date {\tt apec} model, to maximise the fitting
  speed. The only variable parameters are the column density of the
{\tt wabs} component, the plasma temperature of the {\tt mekal}
component (initial value kT = 0.5 keV; range: 0.08 - 20 keV), and its
normalisation. As a consequence, this model is not always an
acceptable fit for these kind of sources, but it is better, in terms
of {\tt goodness} (see Sect.\ref{good}), than a power-law model in
most cases.

Similarly to the absorbed power-law model fit, the plasma temperature
is computed by removing the absorption component and fitting this
simple thermal model in the {\it Soft} band. The resulting temperature
is the one used to fix this parameter in the case it cannot be
constrained. In the case the number of counts in the {\it Soft} band
is $<$ 20 counts, a value of kT = 1 keV is used instead. The fixed
value used in case the column density cannot be constrained is N$_{H}$
= 10$^{20}$ cm$^{-2}$ in all cases.

The spectral-fitting results of this model applied in the {\it Soft}
band are the ones used in the complex model: mekal plus
power-law model.

\subsubsection{Absorbed black-body model}

This model is only applied in the {\it Soft} band to model, for
example, soft emission in AGN, X-ray binaries, and supersoft
novae. The variable parameters are the column density of the {\tt
  wabs} component, and the black-body temperature (initial value kT:
0.5 keV; range: 0.01-10 keV), and its normalisation.

The spectral fitting of this model is carried out in the same way as
the absorbed thermal model. The only difference in this case is that
the fixed value of the temperature in the case of low number of counts
is kT = 0.1 keV.

The spectral-fitting results of this model are used as input
parameters and fixed values in the black-body plus power-law
model fit.

\subsubsection{Absorbed thermal plus power-law model}

This model includes two absorption components: one affecting both the
thermal, and power-law components; and a second one only affecting the
power law component. In the case of AGN for example, this model could
represent the host-galaxy soft emission (thermal component), and the
more absorbed intrinsic AGN emission (power-law component), although it
could also model appropriately emission from X-ray binaries.

The initial parameters are extracted from the absorbed thermal model
fitting in the {\it Soft} band, and the absorbed power-law model
fitting in the {\it Hard} band. If the number of counts in either of
those bands is lower than 50 counts and thus, either of these models
has not been fitted, this complex model is not fitted either. Taking
into account the limit of 500 counts to apply complex modes, it is
reasonable to assume that spectra that lack enough counts in the {\it
  Hard}/{\it Soft} band do not need an additional power-law/thermal
component to model the spectral shape. 

\subsubsection{Double power-law  model}

As the previous model, this model also includes two absorption
components. The photon indices of the two power-law components are not
fixed together so this model could represent several physical
scenarios: an absorbed power-law plus a scattered component; a partial
covering absorber affecting intrinsic power-law emission; and even a
hard power-law plus a soft thermal component if the data at low
energies are of low quality. In this case, the initial values and
fixed parameters are extracted from the absorbed power-law model
fittings in the {\it Soft}, and {\it Hard} bands. As for the previous
model, if either of these fits has not been performed, this complex
model fit is not carried out.

\subsubsection{Absorbed black-body plus power-law model}
\label{bbpo}

In this case there is only one absorption component covering both the
black-body and the power-law components. The black-body component is
often used to phenomenologically represent soft emission in AGN, but
it can also account for disk emission in X-ray binaries. The initial and
fixed values for the parameters are extracted from the absorbed
black-body model fit, and the absorbed power-law model fit in the {\it
  Hard} band.

\subsection{Fluxes}

For each model applied, the observed flux and its errors (at the 90\%
confidence level) are computed in the band used in the spectral
fit. In the case of multiple instrument spectra being jointly fitted,
the flux included in the database corresponds to the average of all
available instruments and exposures for that observation. Fluxes and
errors are reported in erg cm$^{-2}$ s$^{-1}$. In a small number of
cases ($<$ 1\%), {\tt Xspec} fails to compute the errors on the
fluxes. In those cases, only fluxes values are reported in the
database.

\begin{figure}[!htbp]
 \centering
    \includegraphics[angle=-90,width=8cm]{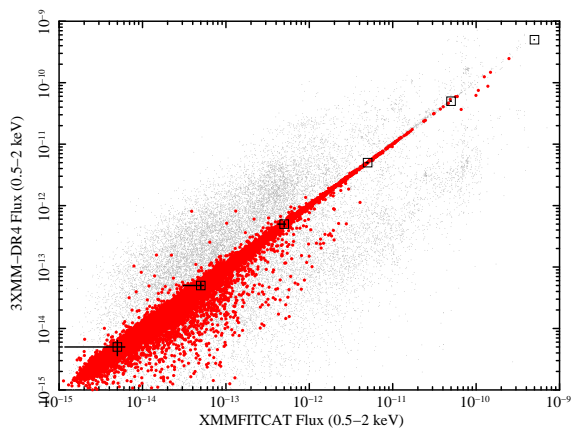}
  \caption{Observed fluxes in the {\it Soft} band (in erg cm$^{-2}$
    s$^{-1}$) from the automated fits against the ones reported in the
    3XMM-DR4 catalogue. Filled circles correspond to consistent fluxes
    between both catalogues, whereas small points correspond to
    non-consistent ones. Square points with error bars represent
    average error sizes at each flux interval.}
\label{fluxsfig}
\end{figure}

A direct comparison between fluxes reported in this database and
within the 3XMM-DR4 catalogue is not possible for all energy bands
(see Table~\ref{bands}), but only in the database {\it Soft} band. The
observed fluxes in the {\it Soft} band obtained from the automated
fits are plotted against the one reported in 3XMM-DR4 in
Fig.~\ref{fluxsfig}. Values in both catalogues agree in 70\% of
cases. Significant differences between both values correspond to one
of the following preferred models: black-body or thermal model; and
power-law models with steep photon indices. The larger number of
consistent fluxes below the on-to-one line is due to larger errors in
the XMMFITCAT fluxes computed in the {\it Soft} band, sometimes
consistent with zero at the 90\% confidence level.

\subsection{Goodness of fit}
\label{good}

C-stat does not provide goodness of fit. As an estimate, the {\tt
  Xspec} command {\tt goodness} is used. This command performs a
number of simulations, 1000 simulations in the case of this database,
and returns the percentage of simulations that gives a lower value of
the statistic. For large return values of this command, the model can
be rejected at the {\tt goodness} value confidence level.

As another proxy of the goodness of fit, the reduced $\chi^2$ value,
after C-stat fitting, is also computed and included in the
database. There is not direct correspondence between {\tt goodness}
and reduced $\chi^2$ values computed this way, although low {\tt
  goodness} values ($<$ 50\%) correspond to low reduced $\chi^2$
values ($<$ 1.5) in 97\% of cases. As it is implemented in {\tt
  Xspec}, C-stat/d.o.f values tend to reduced $\chi^2$ values for
large number of counts. However, most observations (80\%) within the
3XMM-DR4 catalogue have $<$ 1000 counts. A conservative approach is
adopted in the database to decide if a spectral-fit is an acceptable
fit or not. Instead of using the $\chi^2$ values from the C-stat fits,
we used only the value of {\tt goodness}. We consider a fit as an
acceptable fit if the return value of {\tt goodness} is lower than
50\%. Although both estimates behave in a similar way, i.e. they are
similarly ``good'' in separating acceptable from unacceptable fits, by
using {\tt goodness} a smaller number of unacceptable fits are included
within our acceptable criteria (see Sect.\ref{gtest}).
 
As a guide to the user, the simplest model applied in the {\it Full}
band with the lowest value of {\tt goodness} is considered as the
preferred model for that observation. However, since both {\tt
  goodness} and reduced $\chi^2$ values are provided within the
database, the user could decide between both estimates to select the
best-fit model.

\section{Database overview}
\label{overview}
\begin{figure}[!htbp]
 \centering
    \includegraphics[angle=-90,width=8cm]{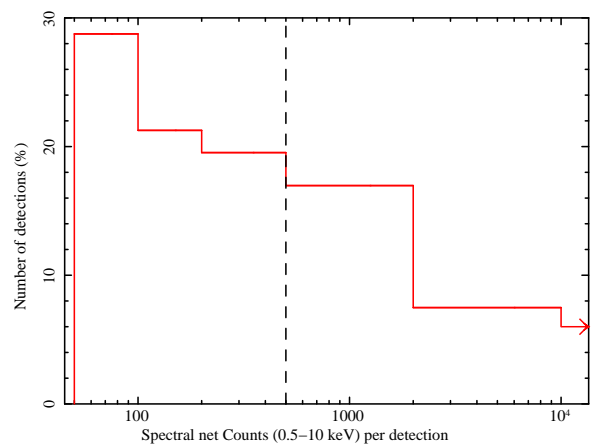}
  \caption{Distribution of net source counts per detection in XMMFITCAT. }
\label{countsdistb}
\end{figure}

\begin{figure}[!htbp]
 \centering
    \includegraphics[angle=0,width=8cm]{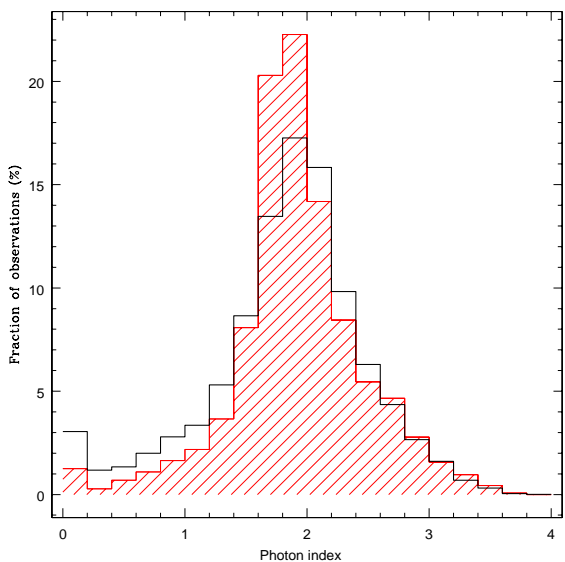}
  \caption{Photon index distribution for the XMMFITCAT detections for which an absorbed power-law model is the preferred model. Empty histogram corresponds to detections with $<$ 500 counts, and line-shaded histogram to detections with $>$ 500 counts. }
\label{podistb}
\end{figure}

The final {\it XMM-Newton} spectral-fit database contains
spectral-fitting results for 114~166 observations, corresponding to
77~954 unique sources. Acceptable fits are found for 90\% of the
observations with less than 500 counts (70\% of all the observations),
whereas they are found for 80\% of the observations with more than 500
counts. This is a remarkable result given the limited number of
spectral models used, and the difficulty often found to obtain
acceptable fits in the case of high number of counts even in manual
fits. The distribution of counts in the XMMFITCAT is shown in
Fig.~\ref{countsdistb}. The vertical line correspond to the limit of
500 counts above which complex models are applied. The distribution of
photon indices (excluding detections for which it had to be fixed) for
the sources best-fitted by an absorbed power-law model is plotted in
Fig.\ref{podistb}. As expected, since most X-ray sources are likely
AGN, the distribution is very similar to the one usually obtained from
X-ray analyses of AGN. We separated observations with fewer than 500,
and more than 500 counts and found and average $\langle\Gamma\rangle$ =
1.8$^{+0.4}_{-0.3}$ and a standard deviation of 0.64, and
$\langle\Gamma\rangle$ = 1.9$^{+0.2}_{-0.2}$ and a standard deviation of
0.54, respectively. Both distributions are very similar, the lower
value for the average $\Gamma$ for detections with $<$ 500 counts
likely due to the higher tail towards lower photon index values.

\begin{figure*}[!htbp]
 \centering
$$
\begin{array}{cc}
\includegraphics[angle=0,width=8cm]{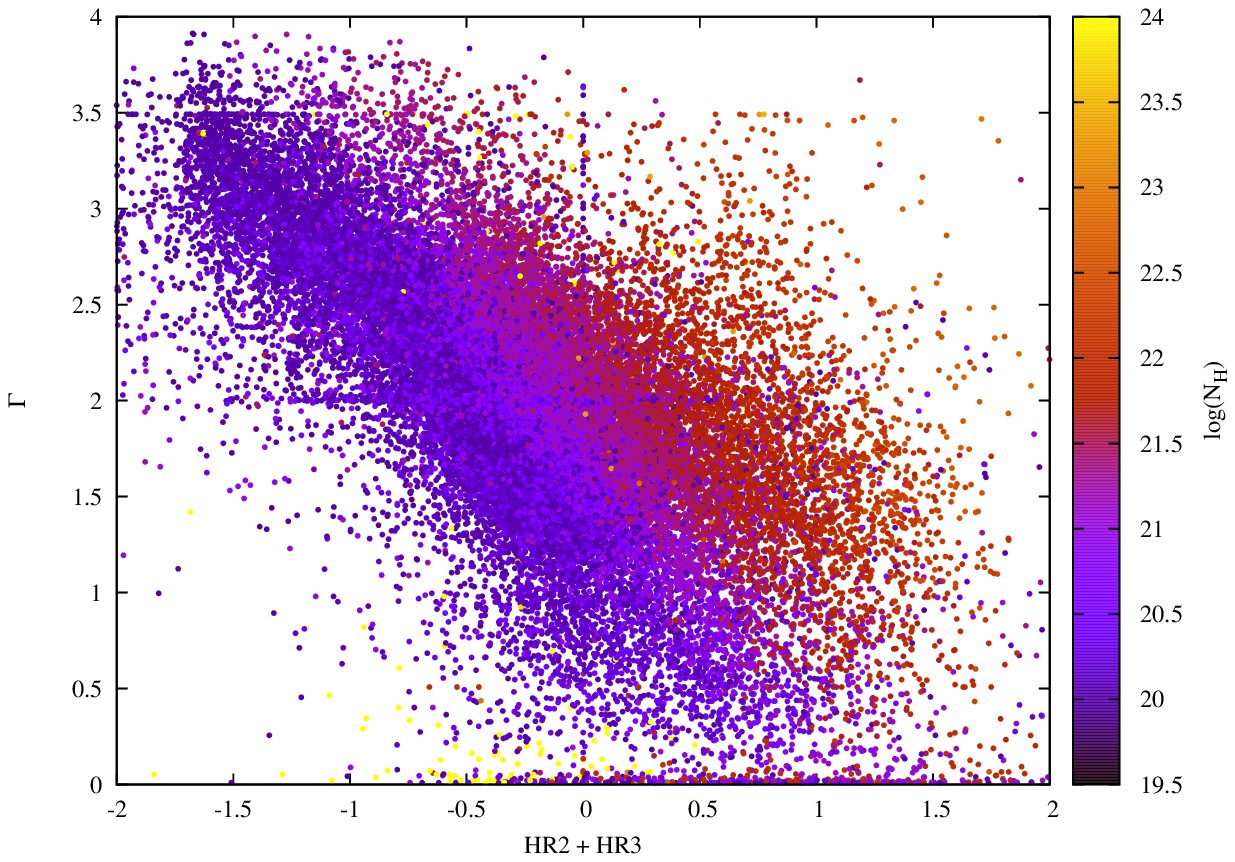} & \includegraphics[angle=0,width=8cm]{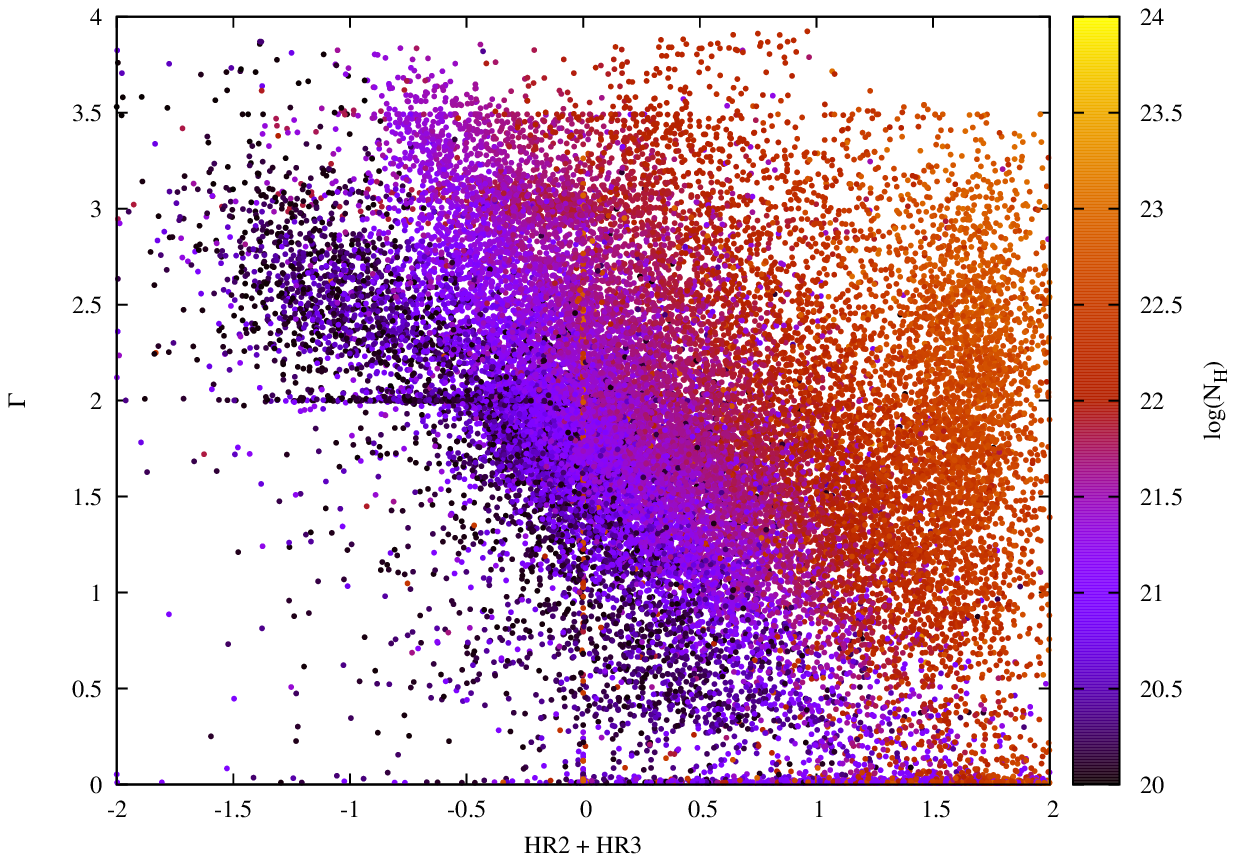}\\
\end{array}
$$
  \caption{Hardness ratios from 3XMM-DR4 against best-fit parameters (photon index and column density) from XMMFITCAT. Left panel: High Galactic latitude sample (|b| > 20). Right panel: Low Galactic latitude sample (|b| < 20).}
\label{hr}
\end{figure*}

Hardness ratios (or X-colors) are used as a proxy of the spectral
shape, and to estimate the column density. Following the band
numbering in Table~\ref{bands}, hardness ratios are defined within
3XMM-DR4 as:
\begin{equation}
$$
{\rm HR_{n}}=\frac{{\rm CR(band\,n+1) - CR(band\,n)}}{{\rm CR(band\,n+1) + CR(band\,n)}}\\
$$ 
\end{equation} where CR(band n) is the count rate in the band
number n. Therefore, a correlation is expected between hardness ratio
values and spectral parameters. To check the consistency between the
results in 3XMM-DR4 and XMMFITCAT, we compared the HR2+HR3 values from
3XMM-DR4, against the best-fit parameters from the absorbed power-law
fit in XMMFITCAT (see Fig.~\ref{hr}). As it can be seen in
Fig.~\ref{hr}, we find a very good agreement between both catalogues.

Less than 1\% of the observations in XMMFITCAT are affected by {\tt
  Xspec} errors. This means that these observations lack spectral
fitting results for one or more, but not all, of the spectral models
that were applied according to their spectral counts. These errors
occur when the model is a very bad representation of the spectral
shape and/or the allowed ranges for the parameters did not encompass
the best-fit values. As a consequence, {\tt Xspec} fails in finding a
minimum and/or falls into an infinite loop.

More than 18\% of the sources within XMMFITCAT have multiple
observations with spectra available. This represents an incredible
source of information for spectral variability studies, that are
usually observationally expensive. 

This database has been already been successfully used to devise a
selection technique for highly absorbed AGN \citep{corral14}. That
work was envisioned as a test of the capabilities of XMMFITCAT in
constructing representative samples of different X-ray sources. We
used the automated spectral-fitting results as a starting point from
which to pinpoint candidate sources, and then confirmed their obscured
nature by using manual fits. We derived an efficiency of our automated
method of $\sim$80\% in selecting highly absorbed AGN.

The database, and the list of spectra used in the spectral fits, can
be retrieved in FITS format from the database project webpage:
http://xraygroup.astro.noa.gr/Webpage-prodec/index.html. The
spectral-fitting results can be also be queried by accessing the
LEDAS\footnote{http://www.ledas.ac.uk/arnie5/arnie5.php?action=advanced\\\&catname=3xmmspectral}
(LEicester Database and Archive Service), and the
XCAT-DB\footnote{http://xcatdb.unistra.fr/3xmm/}, that also includes a
data visualisation tool.

The verification tests carried out, as well as a description of the
database columns can be found in the Appendix.

\begin{acknowledgements}
We thank the anonymous referee for providing us with constructive
comments and suggestions. A. Corral acknowledges financial support by
the European Space Agency (ESA) under the PRODEX program.
\end{acknowledgements}

\bibliographystyle{aa}
\bibliography{XMMFITCAT_ACorral}

\begin{appendix}
\section{Verification procedures}
\label{verification}
In the following sections of the appendix we describe some of the quality
verification tests that have been carried out during the construction
of the database.

\subsection{Goodness of fit: C-stat versus $\chi^2$ fitting}
\label{gtest}
$\chi^2$ fitting has the advantage of providing goodness of fit,
reduced values ($\chi^2$/d.o.f.) $\sim$ 1 indicating that the model is
a good representation of the data. However, a relatively high number
of counts is needed in order to use this statistic. C-stat fitting
works very well down to 40 spectral counts, but it does not provide
goodness of fit. Two possible solutions to this problem are: to use the
command {\tt goodness} in {\tt Xspec}; and to use C-stat as fitting
statistic but $\chi^2$ as the test statistic (see Sect.\ref{good}).

\begin{figure}[!htbp]
 \centering
 \includegraphics[angle=-90,width=8cm]{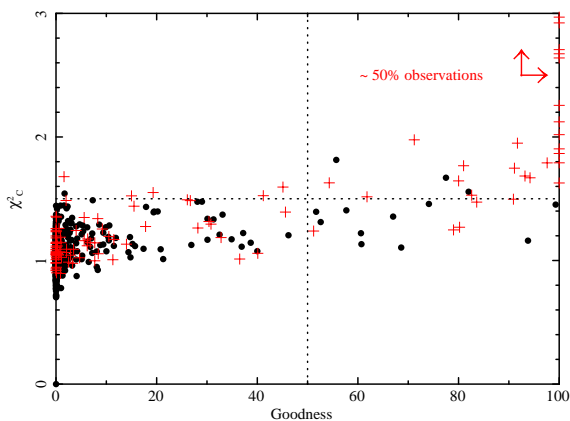}
  \caption{Goodness values versus reduced $\chi^{2}$ values for the
    same spectral fits. Circles and crosses correspond to acceptable
    and unacceptable fits, respectively, according to $\chi^{2}$
    fitting.}
\label{compgood}
\end{figure}

To compare both goodness estimates to define acceptable fits in the
database, a random sample ($\sim$ 2000 detections with more than 200
net counts) was selected from the 3XMM-DR4 catalogue. The automated
fitting procedure was applied both using C-stat fitting to spectra
binned to 1 count/bin, and $\chi^2$ fitting to spectra binned to 20
count/bin. In the second case, $\chi^2$ values are true
representations of the goodness of fit. In Fig.~\ref{compgood}, the
values of {\tt goodness} are plotted against derived $\chi^2$ values
from C-stat fitting ($\chi^{2}_{C}$). Different symbols represent
acceptable (filled circles), and unacceptable (crosses) fits from
the $\chi^2$ fitting of the same observations. We define as an
acceptable fit a $\chi^2$ fit with a reduced $\chi^2$ value $<$
1.5. The arrows on the plot indicate that 50\% of the unacceptable fits
according to $\chi^{2}$ fitting lie above $\chi^{2}_{C}$ > 3 and {\tt
  goodness} > 80. We find that there are no limits for {\tt goodness}
or $\chi^{2}_{C}$ that allow us to distinguish unambiguously between
acceptable and unacceptable fits by using these estimates. The
compromise adopted in the database is to define an acceptable fit as a
fit with a goodness value $<$ 50\%, that roughly corresponds to
reduced $\chi^{2}_{C}$ < 1.5, but classify less probably-bad fits as
acceptable fits.

It is important to note that both estimates, {\tt goodness} < 50\% and
$\chi^{2}_{C}$ < 1.5, distinguish between acceptable and unacceptable
fits (as defined by $\chi^{2}$ < 1.5, or > 1.5, respectively),
similarly well. We find that both criteria classify correctly 90\% of
the spectral fits. The disadvantage of each criterion is to include
more bad fits within the acceptable fits in the case of
$\chi^{2}_{C}$, and classifying more good fits as unacceptable, in the
case of {\tt goodness}.

For lower number of counts, the comparison is more difficult since
$\chi^{2}$ fitting is less reliable. To compare the reliability of both
estimates we made use of simulations. We simulated $\sim$ 4000
observations with fewer than 200 counts, using an absorbed power-law
and an absorbed thermal model, and considering a wide range of values
for the photon index, the thermal component temperature, and the
column density. Assuming that power-law simulated models should be
well-fitted by a power-law, we find again that using {\tt goodness} we
miss a larger number of good fits than using $\chi^{2}_{C}$ (17\% fits
are considered unacceptable according to {\tt goodness} values,
whereas only 2\% are according to $\chi^{2}_{C}$ values). However,
simulated thermal models are classified as acceptable fitted by a
power-law model according to $\chi^{2}_{C}$ in 70\% of the cases,
whereas they are only classified as acceptable according to {\tt
  goodness} in 50\% of them. Both criteria seem to classify correctly
the same fraction of spectral fits ($\sim$ 80\%, lower than the 90\%
for more than 200 counts), but again using {\tt goodness} we miss more
good fits, and using $\chi^{2}_{C}$ we miss-classify more bad fits as
good fits.

Although it is not the purpose of this database to decide between
models but to provide as much information as possible about the
spectral shape, the {\tt goodness} values are used, as a guide to the
user, to both distinguish between acceptable and unacceptable fits,
and to select the best-fit model within the database. However, {\tt
  goodness} and $\chi^{2}_{C}$ values are both included in the
database so the users may consider either or both values to define
their own acceptable/unacceptable classification or to select the
best-fit model.

\subsection{Dependence on source type}
To study if the automated preferred models and best-fit parameters are
in agreement with what is often found from manual spectral analyses,
we constructed a sample of $\sim$ 500 sources including stars
(2XMM/Tycho sample, \citealt{pye08}), Low Mass X-ray binaries, (LMXB,
from the catalogue in \citealt{liu07}), High Mass X-ray Binaries
(HMXB, from the catalogue in \citealt{liu06}), normal galaxies, and
Active Galactic Nuclei (AGN from the XMM-Newton Bright Sample, XBS,
\citealt{ceca04}). These kinds of sources are representative of the
most common types expected to be found within the XMM-Newton
catalogue. This test sample also includes a great variety in spectral
quality, including bright targeted sources as well as serendipitous
much more fainter sources.

We then applied the automated spectral-fitting process to the 500
sources. In 6\% of cases an acceptable fit was not found (goodness of
fit larger than 50\% for all models). All these cases correspond to
X-ray binaries and stars with large numbers of net counts ($>$ 15000),
and a very complex spectral shape, and for which a much more detailed
spectral analysis has already been published in the literature. Most
sources with less than 1000 counts are well-fitted by using simple
models. But it is also important to note that a good fit is also found
for 20\% of the sources with more than 15000 counts.

\begin{figure}[!htbp]
 \centering
    \includegraphics[angle=0,width=8cm]{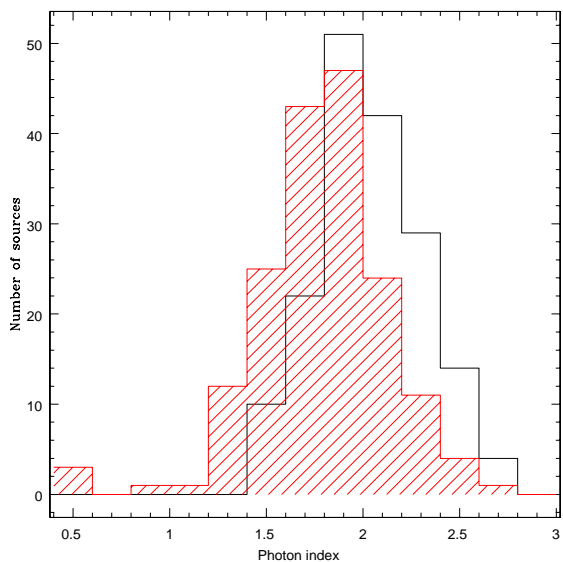}
  \caption{Photon index distribution for XBS AGN obtained from the automated spectral fit (line-shaded histogram) and from the manual fit in \cite{corral11}(empty histogram).}
\label{gdistb}
\end{figure}

\begin{figure}[!htbp]
 \centering
    \includegraphics[angle=0,width=8cm]{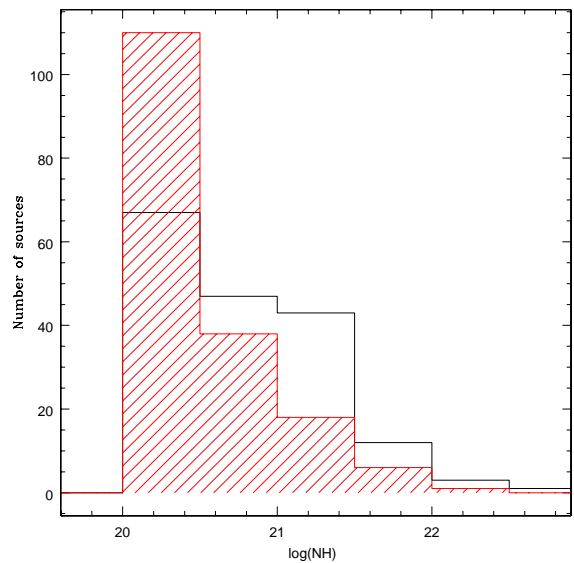}
  \caption{Column density distribution for XBS AGN obtained from the automated spectral fit (line-shaded histogram) and from the manual fit in \cite{corral11}(empty histogram).}
\label{nhdistb}
\end{figure}

The large majority of AGN (95\%) are well-fitted by using a simple
absorbed power-law model. The photon index distribution is in good
agreement with published results from manual spectral analyses of
X-ray selected AGN. A manual spectral fitting analysis of the same
sample of AGN was presented in \citet{corral11}, which has been used
to compare our results. The individual values for the photon index
from the automated fits are systematically lower than the ones
presented in \citet{corral11} (see Fig.~\ref{gdistb}), but consistent
within errors in most cases. The same applies for the column density
values (see Fig.\ref{nhdistb}), the values in \citet{corral11} being
higher likely because of the also higher values of the photon indices
and the effect of redshift, but consistent with the automated ones
within errors. These small differences are likely caused by one or
more of the following intrinsic differences between both analysis:
\begin{itemize}
\item{{\it Different energy channels taken into account:} The fitting
    statistics used in \citet{corral11} was $\chi^{2}$ instead of
    C-stat. One of the main differences between the two statistics is
    that spectral channels were added together to contain at least a
    minimum number of counts in \citet{corral11}. This can lead to the
    removal of high energy spectral channels (usually less populated)
    and as a consequence, the lost of counts at high energies, whereas
    using C-stat almost all detected photons in the energy range in use are
    taken into account. These high energy channels usually contain
    only a small number of counts, so they are often removed if only
    added channels with a minimum number of counts are considered,
    which can result in a softer spectrum used during the spectral
    fit, and a resulting higher value for the photon index.}
\item{{\it Different fitting statistics:} The systematic differences
  can be produced by the use of C-stat instead of $\chi^2$ too, and it
  could be simply due to the fact that C-stat fitting is better in the
  low-count mode. In fact, the higher the number of counts, the closer
  the parameter values become between C-stat and $\chi^2$ fitting.}
 \item{{\it Different spectral-fitting methods:} It is also important
   to note that, unlike previously reported manual analysis of these
   samples, the automated fits do not make use of any assumptions
   regarding the type of source under consideration, nor do they include
   information about the source redshifts. Besides, very hard photon
   indices ( $<$ 1.4) were not allowed in \citet{corral11}. If a very
   hard photon index was found, either it was fixed to 1.9, or
   additional spectral components were added to the fit.}
\end{itemize}

The results for the rest of source types are also in agreement with
published results. Most stars, normal galaxies, and LMXBs are
better-fitted by using soft models, i.e. by power law models with a
steep (most of them $>$2) photon index, or by thermal models. HMXBs are
better-fitted by hard models, i.e. by models including a power-law
component with a flat photon index (most of them $<$2). 

\subsubsection{Manual testing}
We constructed another randomly selected sample of 500 sources,
extracted from 3XMM-DR4, in order to manually check the spectral
results. The strategy was to apply the automated spectral fitting
procedure to these sources and then, to fit them also manually by
using the same set of models and compare the results. The
selected sample spans a wide range in spectral quality very similar to
the one spanned by the full XMMFITCAT.

\begin{figure*}[!htbp]
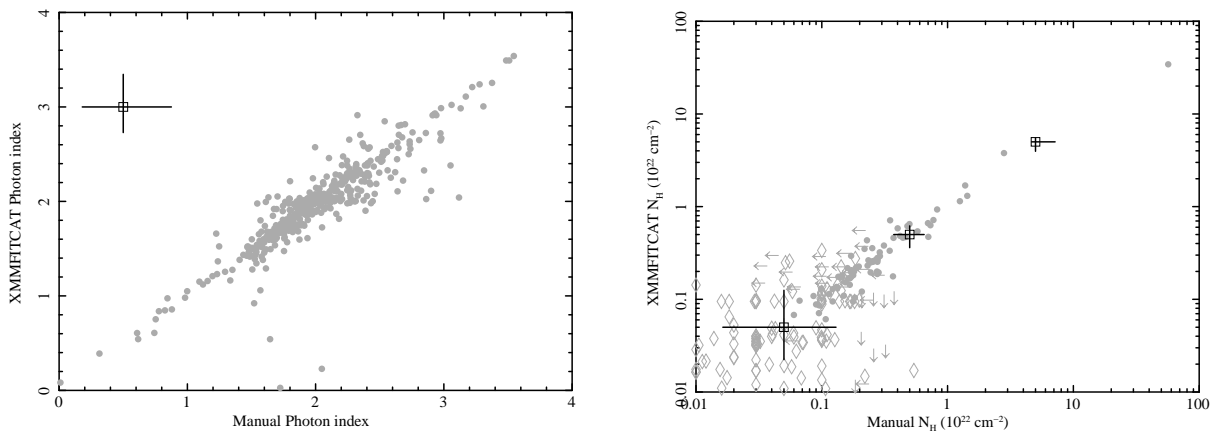

 \centering
$$
\begin{array}{cc}
\includegraphics[angle=-90,width=8cm]{figurea4a.ps} & \includegraphics[angle=-90,width=8cm]{figurea4b.ps}\\
\end{array}
$$
  \caption{Best-fit parameters (left: photon index, right: absorbing
    column density) from the manual fits compared to the ones obtained
    from the automated fits. Empty symbols and arrows on the right
    panel correspond to upper limits. Square points with errors
    represent average error sizes. }
\label{testing2}
\end{figure*}

As a first step, we compared the values of the resulting spectral
parameters, such as the power-law photon index, the temperature of the
thermal component, or the inferred flux. We find an excellent
agreement between these values in almost all cases. Significant
deviations between values derived from different methods only occur if
the model is not an acceptable fit. The values obtained from the
automated fits against the ones obtained manually, for the photon
index and the absorbing column density, are plotted in
Fig.~\ref{testing2}. Note that the most significant differences,
although consistent within errors, for the column density values occur
only for low values of this parameter, i.e., when the absorption
component does not affect significantly the spectral shape.

\begin{figure}[!htbp]
 \centering
\includegraphics[angle=-90,width=8cm]{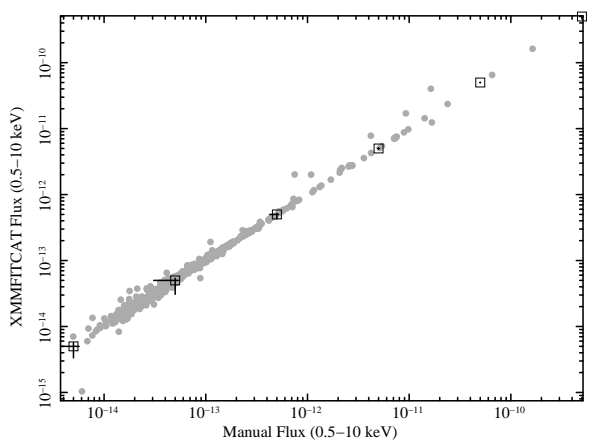}\\
  \caption{Fluxes (in erg cm$^{-2}$ s$^{-1}$) computed by using manual fits versus the ones obtained by the automated fits. Square points with errors correspond to the average error at each flux.}
\label{testing1}
\end{figure}

We also checked if the model considered as our best-fit model was the
same for the manual and automated fits. For the one-component models,
the model selected as our best-fit model by the automated process was
the same as for the manual process in 95\% of the cases. In the case
of two-components models, it is extremely difficult to decide between
two acceptable models even if we could take into consideration the
source type. Nevertheless, the manually derived spectral parameters
are in agreement with the ones obtained from the automated fits in
almost all cases (see, for example, the computed fluxes plotted in
Fig.~\ref{testing1}).

\subsubsection{Simulated data}

Finally, we tested the ability of the automated procedure to
distinguish between spectral models, and its accuracy at retrieving
the intrinsic shape. To this end, we selected yet another random
sample from 3XMM-DR4 of 1000 detections with the same count
distribution as the full XMMFITCAT. Then, we used the preferred model
and parameters to simulate 10 times each source and background
spectra, and applied the automated procedure to the simulated data.

\begin{figure}[!htbp]
 \centering
\includegraphics[angle=-90,width=8cm]{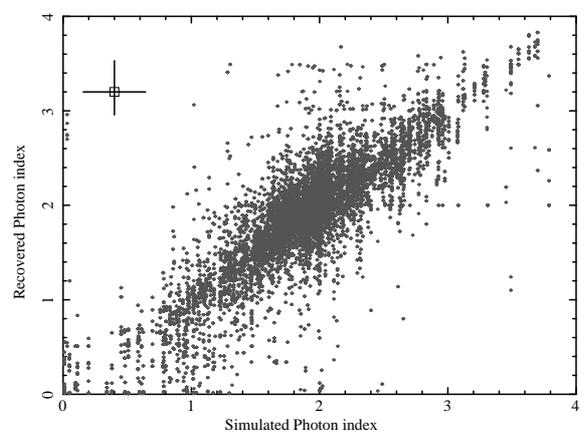}\\
  \caption{Simulated photon indices versus the ones obtained from the automated spectral fits. Square point with error bar represents the average error size.}
\label{compp}
\end{figure}

We find that the simulated model and the preferred model after the
automated fits agree in $\sim$ 87\% of the cases. Nevertheless, the
preferred model is given as a guide to the user, and it is not the aim
of this database to distinguish between models, but to provide a good
representation of the spectral shape. We find that the vast
  majority of the best-fit parameters from the automated fits are
  consistent within errors with the input parameters of the
  simulations. Therefore, the automated procedure is very successful
  in recovering the simulated spectral shape. As an
example, the simulated photon indices are plotted against the ones
obtained after the automated fits in Fig.~\ref{compp}.

\section{Database columns}

 The catalogue contains 214 columns. A description of each column is
 given in the following sections. The name is given in capital
 letters, the FITS data format in brackets, and the unit in square
 brackets. For easier reference the columns are grouped into five
 sections. Non-available data are represented by a -99 value within
 the FITS table.
   
\subsection{Identification of the detection}
The first eight columns of the database are also contained within the
3XMM-DR4 catalogue, and their purpose is to identify each source
detection and spectral products, and to this end, they share the same
column name, format, and values as in 3XMM-DR4.\\*\\*
{\bf IAUNAME} (21A): the IAU name assigned to the unique SRCID.\\ 
{\bf DETID} (J): a consecutive number which identifies each entry
(detection) in the catalogue. The DETID numbering assignments in
3XMM-DR4 bear no relation to those in 2XMMi-DR3 but the DETID of the
nearest matching detection from the 2XMMi-DR3 catalogue to the
3XMM-DR4 detection is provided via the DR3DETID column (not included
in the XMMFITCAT table) within 3XMM-DR4. \\ 
{\bf SRCID} (J): A unique number assigned to a group of catalogue
entries which are assumed to be the same source. The process of
grouping detections in to unique sources has changed since the 2XMM
catalogue series. The SRCID assignments in 3XMM-DR4 bear no relation
to those in 2XMMi-DR3 but the nearest unique sources from the
2XMMi-DR3 catalogue to the 3XMM-DR4 unique source is provided via the
DR3SRCID column (not included in XMMFITCAT).\\
{\bf OBS\_ID} (10A): The XMM-Newton observation identification.\\
{\bf SRC\_NUM} (J): the (decimal) source number in the individual source
list for this observation as determined during the source fitting
stage; in the hexadecimal system it identifies the source-specific
product files belonging to this detection.\\
{\bf SRC\_HEX} (4A): Source number expressed in the hexadecimal system to
identify the source-specific product files belonging to this
detection.\\
{\bf SC\_RA (D)} [deg]: The mean Right Ascension in degrees (J2000) of all
the detections of the source SRCID weighted by the positional errors.\\
{\bf SC\_DEC (D)} [deg]: The mean Declination in degrees (J2000) of all the detections of the source SRCID weighted by the positional errors.\\
   
\subsection{Spectral source counts}

Give the restriction imposed on the number of counts per individual
instrument spectra, all available spectra for each observation are not
always used in the spectral analysis. The number of counts reported in
the table correspond to the number of counts used during the spectral
fit. This number is computed by adding the number of source
(background subtracted) counts for all instruments and exposures used
in the spectral analysis.\\*\\* 
{\bf T\_COUNTS (D)} [count]: spectral background subtracted counts in
the {\it Full} band (0.5-10 keV) computed by adding all instruments
and exposures for the corresponding observation. A number of counts
equal to -99 means that the number of counts is $<$ 50 counts and the
spectral fit in this band is not performed.\\ 
{\bf H\_COUNTS (D)}
[count]: spectral background subtracted counts in the {\it Hard} band
(2-10 keV) computed by adding all instruments and exposures
for the corresponding observation. A number of counts equal to -99
means that the number of counts is $<$ 50 counts and the spectral fit
in this band is not performed.\\ 
{\bf S\_COUNTS (D)} [count]: spectral
background subtracted counts in the {\it Soft} band (0.5-2 keV)
computed by adding all instruments and exposures for the
corresponding observation. A number of counts equal to -99 means that
the number of counts is $<$ 50 counts and the spectral fit in this
band is not performed.\\

\subsubsection{Galactic column density}
{\bf GNH (D)} [10$^{22}$ cm$^{-2}$]: Galactic column density in the direction of the source from the Leiden/Argentine/Bonn (LAB) Survey of Galactic HI \citep{kalber05}.

\subsection{Spectral model names}
\label{amodels}
Model related columns start with the model name. Model names are as
follows:
\begin{itemize}
\item WAPO: absorbed power-law model applied in the {\it Full} band.
\item WAPOS: absorbed power-law model applied in the {\it Soft} band.
\item WAPOH: absorbed power-law model applied in the {\it Hard} band.
\item WAMEKAL: absorbed thermal model applied in the {\it Full} band.
\item WAMEKALS: absorbed thermal model applied in the {\it Soft} band.
\item WABBS: absorbed black-body model applied in the {\it Soft} band.
\item WAMEKALPO: absorbed thermal plus power-law model applied in the {\it Full} band.
\item WAPOPO: absorbed double power-law model applied in the {\it Full} band.
\item WABBPO: absorbed black-body plus power-law model applied in the {\it Full} band.
\end{itemize}

\subsection{Spectral-fit summary columns}
{\bf A\_FIT} (L): The value is set to True, if an acceptable fit,
i.e. {\tt goodness} value $<$ 50\%, has been found for the models applied in
the {\it Full} band, and to False otherwise.\\
{\bf P\_MODEL} (J): The data preferred model, i.e., the simplest spectral
model with the lowest {\tt goodness} value among the models applied in the
{\it Full} band. Each model is assigned a number as follows:
\begin{enumerate}[start=0]
\item WAPO.
\item WAMEKAL
\item WABBPO
\item WAMEKALPO
\item WAPOPO
\end{enumerate} 

\subsection{Columns containing spectral fitting results}
The remaining columns refer to the spectral-fitting results for the
different models applied. Columns for a single model start with the
model name as listed in Sect.\ref{amodels}.

\subsubsection{Spectral-fit summary for a specific model}
{\bf MODEL\_FIT} (J): spectral-fit summary for the model MODEL. The possible
values for this column correspond to the different situations that may
occur during the spectral fits, and they are described as follows:
\begin{enumerate}[start = 0]
    \item The spectral fit was performed, and the model is considered an acceptable fit, i.e, the value returned by the command {\tt goodness} is lower than 50\%.
    \item The spectral fit was performed, but the value returned by the command {\tt goodness} is greater than 50\%.
    \item The spectral fit was not performed because the number of counts in the {\it Soft} band is lower than 50 counts.
    \item The spectral fit was not performed because the number of counts in the {\it Hard} band is lower than 50 counts.
    \item Complex-model (WAMEKALPO, WABBPO, or WAPOPO) spectral
      fit was not performed because the number of counts in the full
      band is lower than 500 counts.
    \item Complex-model (WAMEKALPO, WABBPO, or WAPOPO) fit results not
      reported because soft model component (thermal or black-body) is
      not significant, i.e., its normalisation is consistent with 0 at
      the 90\% confidence level.
    \item Complex-model (WAMEKALPO, WABBPO, or WAPOPO) fit results not
      reported because hard model component (power-law) is not
      significant, i.e., its normalisation is consistent with 0 at the
      90\% confidence level.
    \item No best-fit parameters found. This may occur if the allowed
      ranges for the parameters to vary and/or the spectral model are
      not a good representation of the data, and {\tt Xspec} falls into an
      infinite loop or fails to find a minimum.
\end{enumerate}

\subsubsection{Parameters and errors}

In the case of absorption components, the column density and errors
reported are in units of 10$^{22}$ cm$^{-2}$. In the case of thermal
and black-body components, the temperatures reported are in
keV. Power-law component normalisation and errors are in photons
keV$^{-1}$ cm$^{-2}$ s$^{-1}$. \\*\\*
{\bf MODEL\_PARAMETER} (D): parameter value. \\*
{\bf MODEL\_PARAMETER\_HI} (D): upper error (at the 90\% confidence
level). In the case of fixed parameters, the error values are equal to
-99.\\*
{\bf MODEL\_PARAMETER\_LO} (D): upper error (at the 90\% confidence
level). In the case of fixed parameters, the error values are equal to
-99.\\*  
{\bf MODEL\_PARAMETER\_ERR} (J): Flag on the parameter error
calculation. The different flags are:
\begin{enumerate}[start = 0]
    \item no errors occurred during the parameter error computation.
    \item parameter error computation hit hard lower/upper limit.
    \item parameter fixed to the value computed in the {\it Soft} band by using a simple model.
    \item parameter fixed to the value computed in the {\it Hard} band by using a simple model.
    \item parameter fixed to its initial value.
\end{enumerate}

\subsubsection{Fluxes}

Reported fluxes are computed in the band used in the spectral fit.\\*\\*
{\bf MODEL\_FLUX} (D) [erg cm$^{-2}$ s$^{-1}$]: the mean observed flux of all instruments and exposures for the corresponding observation, in
the energy band used in the spectral fit.\\*
{\bf MODEL\_FLUX\_HI} (D) [erg cm$^{-2}$ s$^{-1}$]: Flux upper error at the 90\% confidence level. If {\tt Xspec} failed to compute the errors, a -99 value is listed.\\*
{\bf MODEL\_FLUX\_LO} (D) [erg cm$^{-2}$ s$^{-1}$]: Flux lower error at the 90\% confidence level. If {\tt Xspec} failed to compute the errors, a -99 value is listed.

\subsubsection{Fitting statistics}

{\bf MODEL\_CSTAT} (D): C-stat value.\\*
{\bf MODEL\_DOF} (I): Degrees of freedom.\\*
{\bf MODEL\_REDCHI} (D): Reduced $\chi^2$ from the C-stat fitting.\\*
{\bf MODEL\_GOODNESS} (D): Return value of the command {\tt goodness} in {\tt Xspec}.
\end{appendix}
\end{document}